\documentstyle[osa,12pt]{revtex}

\begin{document}
\title{The ordered and orientationally disordered crystalline phases of the
flexible C$_4$F$_8$ molecule}
\author{Z. Gamba}
\address{Dept. of Physics, CNEA, Av. Libertador 8250, (1429) Buenos Aires,
 Argentina.
E-mail: gamba@cnea.gov.ar}
\author{and}
\author{B. M. Powell}
\address{National Research Council of Canada, Steacie Institute for Molecular\\
Sciences, Neutron Program for Materials Research, Chalk River Labs., Chalk\\
River, Ontario, K0J 1J0 Canada. E-mail: qrtrbell@webhart.net}
\maketitle

\begin{abstract}
There is ample experimental evidence for the existence of several crystalline
phases of C$_4$F$_8$, although they still have been not clearly identified.
In this paper we report a series of molecular dynamics (MD) simulations
using a partially flexible molecular model, which takes into account the
mixing of the low frequency intramolecular modes and lattice modes.  The
calculations are carried out at constant pressure and constant temperature
and the algorithm employed allows for volume and symmetry changes of
the MD\ sample as a function of thermodynamic variables. Although several
stable crystalline phases are found, their number is still less than 
determined by experiments.\\
\end{abstract}

{\bf Introduction:}

Per-fluorinated organic compounds are chemically very inert and therefore
their study is of interest in enviromental, medical \cite{general0} and
materials sciences \cite{general1}. Perfluorocyclobutane C$_4$F$_8$, is a
flexible molecule and there is experimental evidence, at ambient pressure,
for the existence of several crystalline 
phase transitions \cite{brian1,nmr1,cp1}. C$_4$F$_8$
crystals melt at 233K and an orientationally disordered phase has been found
between 217K and the melting point. This plastic phase has been measured by
X-rays and neutron powder diffraction methods and the obtained data has been
interpreted as an orientationally disordered crystal. The molecular centers
of mass follow a cubic bcc array, with two molecules per unit cell at
(0,0,0) and (1/2,1/2,1/2), the space group is Im\=3m \cite{brian1}.
 NMR \cite{nmr1} and
heat capacity \cite{cp1} measurements determined the existence of four
solid-solid phase transitions at ambient pressure and at 141, 174, 215 and
217K. Furthermore, in ref.[5] a small hysteresis in the heat
capacity loop was found at about 97K. The low temperature phases have not been
determined, but the scarce experimental data suggested that the phase
transitions are related to the onset of dynamical disorder in the molecules'
orientation and in the intramolecular degrees of freedom \cite{nmr1}.

The only calculation on the condensed phases of this compound is a constant
volume MD simulation of the high temperature plastic phase, using a rigid
model molecule \cite{brian1}, that was helpful to analyze the measured
powder diffraction patterns.

Here we study the phase diagram of C$_4$F$_8$ crystals, as
given by a simple and flexible model. The study is performed
$via$ a series of constant pressure and constant temperature (NPT) 
molecular dynamics (MD) simulations, at several temperatures and
zero pressure. Our calculations reproduce the scarce experimental data 
of the plastic phase and
predict structural and dynamical properties of the low temperature phases.\\

{\bf Intra- and intermolecular potentials:}

C$_4$F$_8$ is a cyclic flexible molecule. The energy associated with
its intramolecular modes is similar to the energy of lattice modes in
crystalline phases, except for the high energy of the stretching modes of
 the molecular bonds, which are considered of constant
length in solid phases \cite{brian1,nmr1}.
 In our MD calculations, the only intramolecular degrees of
freedom taken into account are those associated with vibrational modes at low
frequencies. These modes can mix with the lattice modes and may be 
relevant to the onset of structural phase transitions.

In our simulations, the initial molecular geometry corresponds to the
one determined in the
gas phase \cite{molec1,molec2} and used in the analysis of the high
temperature disordered phase \cite{brian1}. It has to be taken into account
that the experimental geometry was determined with a large deviation, due to
thermal motion. From these measurements it cannot be stated whether the
molecular symmetry is D$_{4h}$ (a planar four C ring) or D$_{2d}$ (a
puckered C ring).

No experimental data on intramolecular forces is available for C$_4$F$_8$.
The vibrational spectra of intramolecular modes have been measured in the
gas, liquid and solid phases and in solid matrices (refs. quoted in ref. 
[3]), but no unequivocal identification of the modes and molecular
geometry was possible. Nevertheless, a clear indication of a low barrier to
torsional ring motion was found.

Here we propose a simple  model molecule, which 
includes a low torsional barrier for
the CCCC ring torsion, $\tau $, and a rocking angle, $\beta $, for the CF$_2$
group (that moves mainly rigidly in a plane perpendicular to the
corresponding CCC plane). By comparison to the very well known data for the
molecule C$_4$H$_8$ \cite{c4h8a}, we propose the following intramolecular
potential model and parameters:

a) The potential for the torsional coordinate $\tau $ (or CCCC angle) is a
double well, of the form

\[
V(\tau )=a-b cos^2(\tau )+c cos^4(\tau ), 
\]
with minimums at cos($\tau _0$)=$\pm \sqrt{\frac{b}{2c}}$=
$\pm cos(12.28^{\circ }$)
and a barrier height $\Delta $V=$a$=$\frac{b^2}{4c}$=$k_{\text{B}}225$K.
The value for $\Delta $V is arbitrary, there is no experimental data on it,
but it was taken low enough (225K is near the melting point) to show
 its possible influence on any of the solid-solid phase transitions.

b) The potential wells for the CCC and FCC angles are harmonic, with force
constants
$f_{CCC}=f_{FCC}= k_{B}$900K= 7.48kJ/mol/rad$^2$. 
These bending angles intervene in the rocking motion of the CF2 group. The
 values of $f_{CCC}$ and $f_{FCC}$ are also arbitrary, 
they imply intramolecular frequencies that are high enough
to avoid mixing with the lattice modes and at the same time explain the
measured molecular geometry as a function of temperature.

The sensitivity of our results to the arbitrary values of barrier height 
and bending force constants are discussed in the section Results (b).

The interatomic angles and distances held constant during the 
simulations are: d$_{CC}$=1.566\AA , d$_{CF}$=1.333\AA , angle 
FCF=109.9$^{\circ }$. 
The initial values for the other quantities are: angle 
CCC=89.34$ ^{\circ }$, angles FCC=110$^{\circ }$ and 117.5$^{\circ }$ and 
torsional angle (CCCC angle) $\tau $=12.28$^{\circ }$. The rocking 
angle, $\beta $, of the CF2 group
is measured as the angle between the bisects of the CCC and FCF 
angles \cite{brian1}.

The intermolecular interactions are taken into account with the atom-atom
Lennard-Jones (LJ) potential model of ref.[3], which does not include 
a set of distributed point charges within the molecule. The LJ parameters
are: $\sigma _{\text{C}}$=3.350\AA , $\sigma _{\text{F}}$=2.825\AA\ , 
$\varepsilon _{\text{C}}$=0.426kJ/mol, $\varepsilon _{\text{F}}$=0.439kJ/mol.
The standard combination rules are applied for cross interactions:
the potential parameters for the interaction between atoms a and b
are calculated as $\sigma _{ab}$=($\sigma _{a}$+$\sigma _{b}$)/2 and
$\varepsilon _{ab}$=
$\sqrt {\text{$\varepsilon _{a}$$\varepsilon _{b}$}}$.
 With these
parameters, the constant volume simulations of ref.[3], performed 
on a statistically  orientationally disordered array
 of rigid puckered molecules,
reproduced the measured X-ray and neutron powder diffraction patterns. The
cut-off radius for the atom-atom interactions is 12\AA\ . Correction terms to
the configurational energy and pressure, due to this finite cut-off, are
taken into account in the usual way: with the integrated contribution of a
uniform distribution of atoms.\\

{\bf Calculations:}

The phase diagram and dynamical properties of C$_4$F$_8,$ at zero pressure,
are studied in the NPT ensemble, by a series of classical constant
pressure - constant temperature MD\ simulations. The implemented 
MD algorithm allows volume and shape fluctuations of the MD sample 
in order to balance the applied
 isotropic external pressure with the internal stresses \cite{algor2}.
This is performed by considering an extended system, which includes as
extra variables the MD box parameters \cite{algor2}. The temperature
 control of the sample follows the approach 
of Nos\'e \cite{algor3,algor4}, which also includes an
external variable. The equations of motion for molecular systems with
internal degrees of freedom are described in ref. [9]. The equations
are integrated $via$ the Verlet algorithm for the atomic displacements 
and the SHAKE\ algorithm for the applied geometrical 
constraints. The 
final MD\ algorithm is identical to that used in a study of black
Newton films \cite{bubbles1}. A further test of the program was
performed by carrying out several calculations at low temperatures and
obtaining the same results as with the program for rigid molecules
MDPOLY, used in ref. [3].

The first series of MD runs consisted of two MD boxes, with an initial array
of 4x4x4 and 6x6x6 cubic bcc unit cells (128 and 432 molecules respectively),
 using the experimental lattice parameters and
disordered molecular orientations, as in 
ref.[3]. These samples were first equilibrated, at 
 P=0kbar and T=224K, for 30000 time
steps (of 0.01ps) with a constant volume algorithm. The 
chosen thermodynamic  parameters are within the small range in
which the plastic phase is found \cite{brian1}.
Afterwards, NPT runs were performed at 220K and the temperature was lowered 
in steps of 25K. At each point of the phase diagram the samples
were equilibrated for 20000 to 30000 time steps (of 0.01ps) and analyzed
 in the following 10ps. Near the phase transitions the equilibration
times had to be increased by a factor of 5-6. Both samples will be called ''cubic
samples'', but this refers only to their initial high temperature structures.

Since these calculations did not show up the expected number of
 phase transitions, a
search for the structure of lowest configurational energy, as given by this
model potential, was performed. The final structure is monoclinic 
 and is described in the following section. Increasing
the temperature of the new sample (hereafter it will be called ''monoclinic
sample'', refering to its initial low temperature ordered structure),
a second series of runs was performed.                 
This procedure is expected to minimize the hysteresis of the possible phase
transitions \cite{mottjones}.

Another sample, initially in trigonal symmetry  was studied upon heating
and is described also in
the following section. This possible structure
was suggested by additional measurements, not reported in ref. [3].\\

{\bf Results:}

{\bf a) Crystalline phases:}

Here we include the results of our calculations in the NPT ensemble of
three samples that gave uniform structures at some point of the phase
diagram. When lowering the temperature of the ''cubic'' sample, or during
the search of the low configurational energy structures, 
many samples showed
twinning or conserved some degree of orientational disorder and were
disregarded. Lengthy calculations were needed to obtain the included
results. A similar problem was found experimentally when trying to identify
the low temperature phases \cite{brian1}.

At 220K the calculated ''cubic'' sample of 432 molecules, remained in
a cubic bcc symmetry, space group Im\=3m, with very large 
fluctuations in the lattice parameters. The averaged thermodynamic 
parameters are T=220(4)K, P=0.0(2)kbar and the lattice
 parameters are a= 6.9(7)\AA\ and angles=90.0(7)deg., which 
compare well with the experimental value of 7.02\AA\
at 224K \cite{brian1}. The calculated pair distribution function and
structure factor, not included here, showed a bcc array for the centers of
mass. The measurement of reorientational times and histograms of 
Euler's angles for the molecules orientations determined that the phase is
plastic. The molecules are dynamically disordered, with the 
molecular planes preferentially oriented  parallel to the cubic
 crystallographic faces, on a time average.
 The disorder is dynamic, with decay times of $\sim $3ps, as obtained
from the orientational time self-correlation functions.

The experimental heat of sublimation is 
$\Delta $H$_{sub}$=28.7kJ/mol at 233K \cite{hsub}. We
estimate $\Delta $H$_{sub}$=30.3kJ/mol at 220K, by adding the entropic
contribution of our calculated vibrational density of states
to the configurational energy \cite{bornhuang}.

At 220K, the "cubic" sample of 128 molecules distorts to a
monoclinic symmetry, instead of remaining cubic. The 6x6x6 unit cells
sample shows this distortion for T$\leq $200K.
These results show that the proposed model molecule reproduces
the plastic phase in large samples, which have
 lower thermal fluctuations of the
lattice parameters. Nevertheless, we believe that our 
high temperature plastic sample is only
marginally the "most stable" structure, large equilibration runs
were necessary, first at constant volume and then at constant pressure,
before being able to obtain the included results.

No further transitions were found for these ''cubic'' samples and this was
the reason to perform a search for the crystalline structure with lowest
configurational energy. Fig. 1 shows this structure, which 
according to our calculations is
the most stable one. Not surprisingly, it is similar to the low
temperature structure of C$_4$Cl$_8$ \cite{c4cl8s}, a monoclinic unit cell
with Z=2 and space group D$_{2h}^2$ (P2$_1$/m). The lattice parameters, at
T=50K, are a= 6.67(2)\AA , b= 4.89(2)\AA , c= 8.42(2)\AA\ and
 $\beta $= 78(1)deg. The two molecules are located at (0,0,0) 
and (1/2,1/2,1/2) and are related by an inversion center at (1/4,1/4,1/4).

A ''monoclinic'' sample of 4x6x4 of the monoclinic unit cells (192 molecules) 
was studied upon
heating. This sample shows a structural phase transition when 
 the temperature is increased
175 to 185K. The high temperature phase is trigonal and  plastic,
as calculated from the pair correlation function for the molecular centers
of mass and reorientational times.

Finally, a "trigonal" sample of 128 molecules was studied upon heating.
This possible symmetry was suggested by preliminary measurements, not
included in ref [3]. The sample initially consisted of 4x4x4 trigonal
unit cells, with two molecules per unit cell, and the molecular planes oriented 
perpendicular to the main diagonal. The lattice parameters of
this initial cell
were a=b=c=7.2\AA\ and angles $\alpha $=$\beta $=$\gamma $=97deg. At
50(1)K the calculated stable structure is triclinic, although the molecular
planes remain parallel. The resulting cell parameters were 
a=5.64(3)\AA, b=7.07(3)\AA, c=6.72(3)\AA\  with angles 
$\alpha $=95.4(3)deg., $\beta $=98.3(3)deg. and $\gamma $=97.1(3)deg.
 For T$\geq $200K this sample shows dynamic orientational disorder.

Fig. 2 shows the configurational energies, the volume per molecule and
estimated free energies of all MD samples as a function of temperature. The
free energy was estimated, at each point of the phase diagram, taking 
into account the contribution of our calculated vibrational
density of states to the entropy \cite{bornhuang}. At low 
temperatures the most stable
sample is the ''monoclinic '' one, at high temperatures all samples have
similar values of configurational and free energies.
 The ''cubic '' sample is slightly more stable, but the 
differences of energy are within the calculated errors.\\

{\bf b) Molecular geometry and dynamics:}

Time averages performed over all molecules show that the average FCC angle
is about 114$^{\circ }$ at all temperatures,
 with deviations of 16$^{\circ }$
at 50K to 23$^{\circ }$ at 225K. These values were obtained from time
averaged histograms of the corresponding angular distributions.

Fig. 3a shows the dependence of the average value of the torsional 
angle \mbox{$<$}$\tau $\mbox{$>$} on temperature. The average 
is over all molecules and over the trajectory in phase space
followed by the system after equilibration.
At 50K $\tau $ is centered at about 7$^{\circ }$, with a very large deviation
 of 4$^{\circ }$ in the "monoclinic" sample, and 6$^{\circ }$ in
the "cubic" sample. At higher temperatures 
\mbox{$<$}$\tau $\mbox{$>$}  decreases 
to 0 and its deviation is
even larger. On a time average the molecules appear nearly planar in all
samples with T$\geq $100K. This change in the dynamics of the 
molecules can, possibly, explain the small
hysteresis measured around 97K \cite{cp1}. Fig. 4 shows the correlation
between the torsional and rocking angles, determined at the lowest temperature
in the ''cubic '' sample. For C$_4$H$_8$ \cite{c4h8a} a relationship of the
form 
\[
\beta =0.21\tau -3.8*10^{-5}\tau ^3. 
\]
was proposed. Fig. 4 shows that, at low temperatures, our measured
correlation could be fitted with a similar function, but our calculations
also show that this correlation is rapidly lost at increased temperatures,
at 100K the distribution is almost flat.

To study the dynamics of the torsional angle $\tau $, several time
correlation functions can be measured. We analyze the 
self-correlation function 
\[
C(t)=\frac 1N\sum_{i=1}^N{\bf A}_i(t).{\bf A}_i(0) 
\]
where N is the total number of molecules and {\bf A }is a vector defined as
 A=0 when the CCCC ring is planar, and A=$\pm 1$ when
 $\tau $=$\pm \tau _0$ , the
 minimum of the potential double well. In our samples at 50K $C(t)$
decays to a final equilibrium value of 0.82, at 100K to 0.40 and at 225K the
function decays exponentially to zero, with a characteristic 
time of $\sim $1ps.  

Clearly that the behavior of \mbox{$<$}$\tau $\mbox{$>$} and 
(\mbox{$<$}$\tau $$^2$\mbox{$>$}-\mbox{$<$}$\tau $\mbox{$>$}$^2$),
    as a function of temperature, depends on the arbitrary value
of the barrier height of the torsional double well   
potential. Since there are no structural changes around 100K in
the three crystalline samples studied here, the change
in molecular dynamics is attributed
 to the heat capacity hysteresis measured at 97K. If
this interpretation is valid, the chosen barrier height 
value is very near the correct one.
The behavior of the rocking angle, as a function of temperature,
 depends on the  $f_{CCC}$ and $f_{FCC}$ values.
Their influence on our calculations 
is indirect. They determine, together with the torsional barrier height, 
the effective molecular volume as
a function of temperature. This fact can be verified with the
calculation of the phase diagram given by a rigid model molecule, using
the same intermolecular potential parameters. The result
is that the order-disorder phase transition, for a
monoclinic sample studied upon heating, is found at T=350K
instead of 185K.\\

{\bf Conclusions:}

The proposed model and potential parameters give good account of
the measured properties of the high temperature plastic phase of 
C$_4$F$_8$, including the packing energy and orientational
preference for the molecular planes to be parallel to the cubic 
faces \cite{brian1}.
Characteristic molecular reorientational times and torsional correlation
decay times are predicted for this phase.

Our calculations also predict a low temperature monoclinic phase, with
orientationally ordered molecules, which is very similar to the room
temperature structure of C$_4$Cl$_8$ \cite{c4cl8s}.

A possible triclinic structure, stable at intermediate temperatures, is
suggested.
 
Our calculations describe the extent of the intramolecular coordinates
disorder. The experimentally observed uncertainty in 
measuring the molecular geometry
is explained by large thermal fluctuations of the torsion and 
rocking angles.
The adequate theoretical model, at temperatures higher than 100K, is of
diffusion between equivalent potential wells with characteristic 
depths of $\sim $k$_B$T. The
expected correlation between the rocking ($\beta $)
 and torsional ($\tau $) angles \cite{brian1} is clearly found 
only at very low 
temperatures. At T$\geq $100K the correlation is lost, both 
angles have an average value of $\sim $0$^{\circ }$ 
with very broad distributions. Above this temperature
 the molecules look almost
planar, as is found experimentally \cite{molec1,molec2}.

The possible contribution of the intramolecular degress of freedom, 
which are dominated by the rocking and torsional angles, to
the observed phase transitions, was 
analysed. The small hysteresis measured around 100K \cite{cp1} 
could be related to the soft double well potential of the
torsional angle.

We could not establish a clear correspondence between
our calculations and all the observed solid-solid structural phase 
transitions. Nevertheless,
several stable structures were found, which can be helpful in 
analysing the neutron and X-rays diffraction patterns \cite{brian1}. 

Our predicted structures and molecular dynamics need further
experimental verifica-

tions. Moreover, the model molecule studied here
is rather simple. A more sophisticated intramolecular potential 
will probably be required to reproduce the complex phase diagram 
of C$_4$F$_8$.\\ 

{\bf Acknowledgement:}
Z. G. thanks CONICET for the grant PIP 0859/98.

\newpage
{\bf Figures:}

Fig. 1: The low temperature monoclinic unit cell, D$_{2h}^2$ (P2$_1$/m), Z=2,
is similar to the low temperature structure of C$_4$Cl$_8$ \cite{c4cl8s}.

Fig. 2: a) Configurational energies, b) volume per molecule and c) estimated
free energies of three MD samples as a function of temperature: ''monoclinic
'' (solid line), ''cubic '' (dashed line) and ''trigonal '' (dotted line).

Fig. 3: Time averaged (a) torsional angle 
, b) rocking angle and c) calculated dispersion of both angles as a function
of temperature. Solid line: ''monoclinic" sample, dashed line: ''cubic '' and
dotted line: ''trigonal ''.

Fig. 4: Correlation between the torsional, $\tau $, and rocking, $\beta $,
 angles, determined at the lowest temperature in the ''cubic '' sample. 


{\bf References:}
\begin{references}
\bibitem{general0}  W. Wyat Gibbs, Scientific American, News and Analysis of
Feb. 1999, p.23.

\bibitem{general1}  A. R. Ravishankara, S. Solomon, A. A. Turnipseed and R.
F. Warren, Science {\bf 259}, 194 (1993).

\bibitem{brian1}  L. S. Bartell and B. M. Powell, Mol. Phys. {\bf 67}, 861
(1989).

\bibitem{nmr1}  E. Szczesniak and J. R. Brookeman, Mol. Phys. {\bf 48}, 1221
(1983).

\bibitem{cp1}  G. T. Furukawa, R. E. McCoskey and M. L. Reilley, J. Res.
Natn. Bur. Stand. {\bf 52}, 11 (1954).

\bibitem{molec1}  H. P. Lemaire and R. L. Livingston, J. Chem. Phys. {\bf 18}%
, 569 (1950).

\bibitem{molec2}  L. S. Bartell and A. Jin, J. Chem. Phys. {\bf 78}, 7159
(1983).

\bibitem{c4h8a}  T. Egawa, T. Fukuyama, S. Yamamoto, F. Takabayashi, H.
Kambara, T. Ueda and K. Kuchitsu, J. Chem. Phys.{\bf \ 86}, 6018 (1987).

\bibitem{algor1}  M. Ferrario and J. Ryckaert, Mol. Phys.{\bf \ 54}, 587
(1985).

\bibitem{algor2}  S. Nos\'e  and M. L. Klein, Mol. Phys. {\bf 50}, 1055 (1983)

\bibitem{algor4}  J. P. Ryckaert and G. Ciccotti, J. Chem. Phys. {\bf 78},
7368 (1983).

\bibitem{algor3}  S. Nos\'e\ , Progress of Teor. Phys. Suppl. {\bf 103}, 1
(1991).

\bibitem{bubbles1}  Z. Gamba, J. Hautman, J. C. Shelley and M. L. Klein,
Langmuir {\bf 8}, 3155 (1992).

\bibitem{mottjones}  N. F. Mott and H. Jones, ''The properties of metals and
alloys'', Dover N. Y., p. 38 (1958).

\bibitem{hsub}  L. S. Batell and Y. Z. Barshad, J. Phys. Chem. {\bf 91},
2893 (1987).

\bibitem{c4cl8s}  T. B. Owen and J. L. Hoard, Acta Cryst. {\bf 4}, 172
(1951).

\bibitem{bornhuang}  M. Born and K. Huang, ''Dynamical theory of crystal
lattices'', Oxford Clarendon Press, 1954, p.39.

\bibitem{c4h8a}  R. Champion, P. D. Godfrey and F. Bettens, J. Mol. Spectr.
{\bf 155}, 18\ (1992).
\end{references}
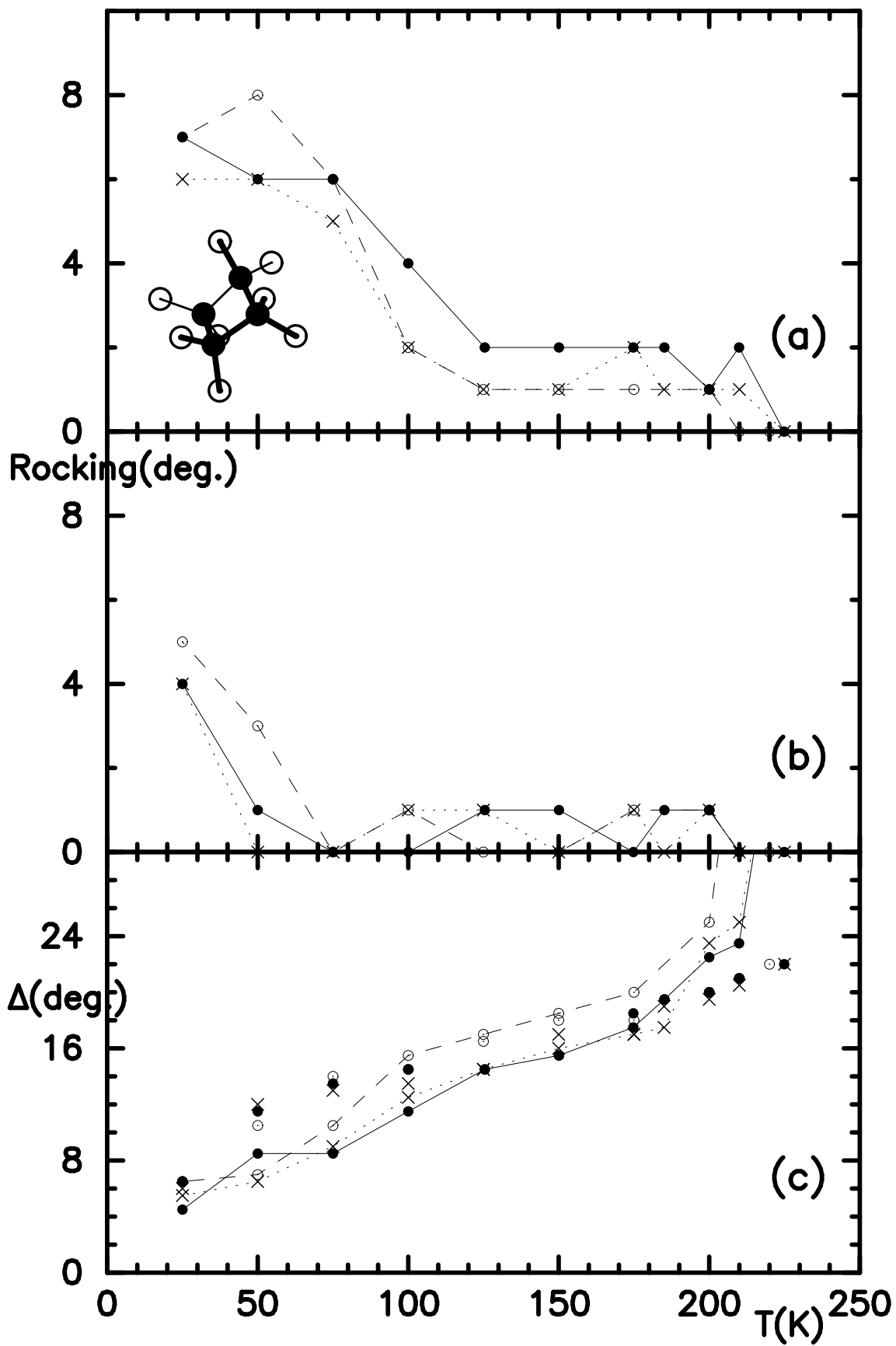
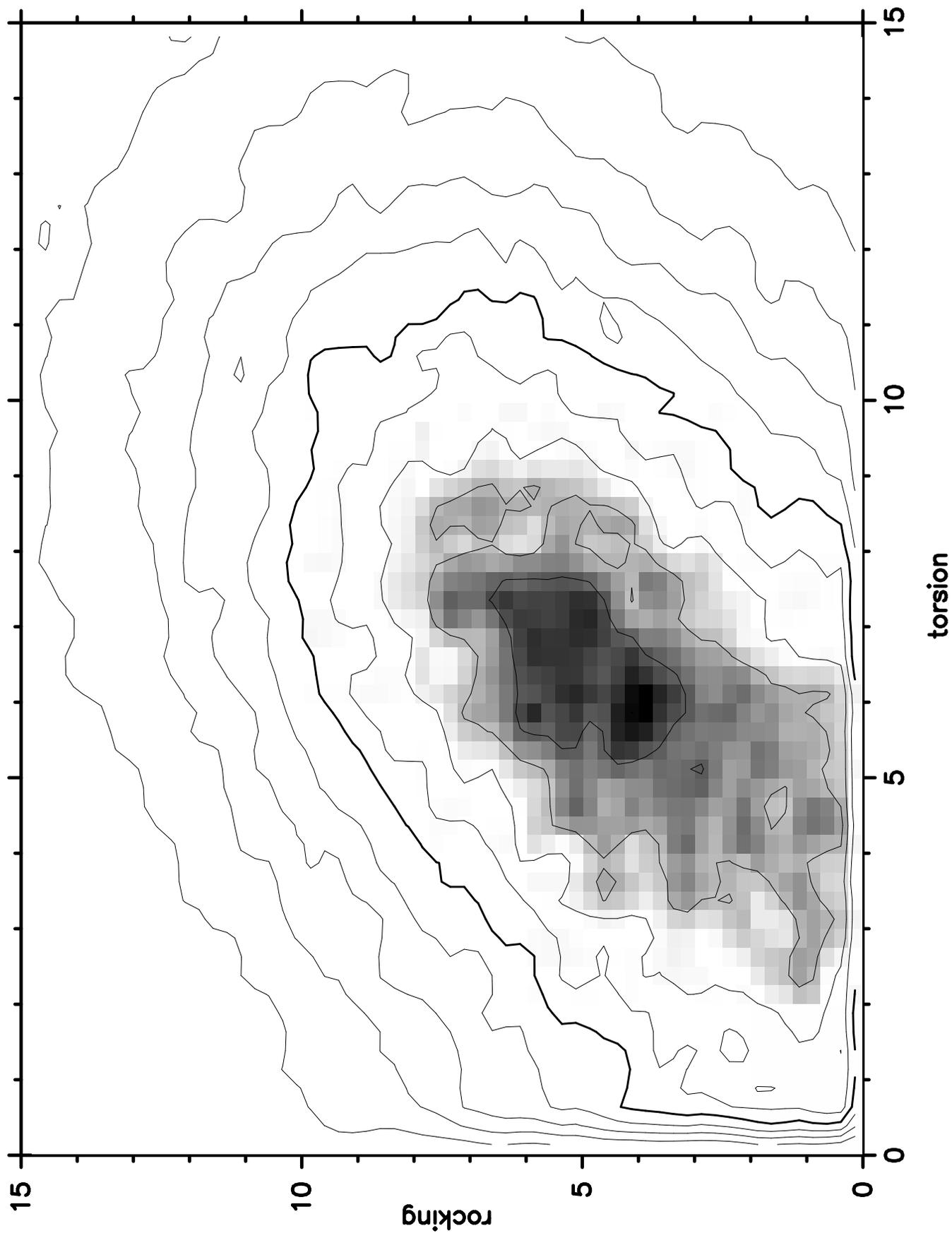
\end{document}